\documentclass[12pt]{article}
\usepackage{epsfig}

\usepackage{axodraw}

\def \gsim{\mathrel{\vcenter
     {\hbox{$>$}\nointerlineskip\hbox{$\sim$}}}}

\author{
Enrico Nardi$^{1,2}$\footnote{enrico.nardi@lnf.infn.it},
 Juan Racker$^3$\footnote{racker@cab.cnea.gov.ar}
and Esteban Roulet$^3$\footnote{roulet@cab.cnea.gov.ar}$\>$    
 \\[5pt] 
$^1${\normalsize
\it INFN, Laboratori Nazionali di Frascati, C.P. 13, I00044 Frascati,
Italy} \\[-2pt] 
$^2${\normalsize \it Instituto de F\'\i sica,
Universidad de Antioquia, A.A.{\it 1226}, Medell\'\i n, Colombia} \\
[-2pt]  $^3${\normalsize
\it CONICET, Centro At\'omico Bariloche, Avenida Bustillo 9500 (8400)
Argentina}}

\title{CP violation in scatterings, three body processes \\ and  the
  Boltzmann equations for leptogenesis}
\begin{document}


\maketitle

\abstract{We obtain the Boltzmann equations for leptogenesis including decay
  and scattering processes with two and three body initial or final states.
   We present an explicit computation  of the
  CP violating scattering asymmetries. We analyze their possible impact in
  leptogenesis, and we discuss the validity of their approximate expressions
  in terms of the decay asymmetry.  In scenarios in which the initial heavy
  neutrino density vanishes, the inclusion of CP asymmetries in scatterings
  can enforce a cancellation between the lepton asymmetry generated at early
  times and the asymmetry produced at later times. We argue that a sizeable
  amount of washout is crucial for spoiling this cancellation, and we show
  that in the regimes in which the washouts are particularly weak, the
  inclusion of CP violation in scatterings yields a reduction in the final
  value of the lepton asymmetry.  In the strong washout regimes the inclusion
  of CP violation in scatterings still leads to a significant enhancement of
  the lepton asymmetry at high temperatures; however, due to the independence
  from the early conditions that is characteristic of these regimes, 
  the final value of the lepton asymmetry remains approximately unchanged.}

\section{Introduction}

Leptogenesis \cite{fu86} provides a very attractive scenario to generate the
baryon asymmetry of the Universe. Indeed, the existence of heavy right handed
neutrinos is strongly motivated by the see-saw mechanism \cite{seesaw}
introduced to generate the observed light neutrino masses.  The Majorana
nature of the right handed neutrinos implies violation of lepton number, the
large values of their masses imply that the out of equilibrium requirement can
hold when they decay, and moreover these decays can be CP violating due to the
phases that are generally present in the Yukawa couplings with the lepton
doublets.  Since the three Sakharov conditions \cite{sa67} can be met, the
question then is a quantitative one, i.e. whether the asymmetry generated in
these scenarios is large enough to account for the observed value.

In recent years, quantitative analyses of leptogenesis  have become
more and more sophisticated, taking into account many subtle but significant
ingredients, such as several washout processes
\cite{lu92,BBP0204-BP9900,pi04-pi05,ha04}, proper subtraction of on-shell
heavy neutrino intermediate states \cite{gi04}, thermal corrections to
particle masses, couplings and decay asymmetries \cite{gi04}, spectator
processes \cite{bu01,na05}, flavor effects 
\cite{ba00,en03-fu05,na06,aba06a,aba06b,
DeSimone:2006dd,Josse-Michaux:2007zj,Shindou:2007se},
and the possible effects of the heaviest right handed Majorana neutrinos
$N_{2,3}$ \cite{ba00,diba05,vi05,eng06} (for reviews of the most recent
results see \cite{review}).

The aim of this work is to discuss in some detail the inclusion of CP
violation in scatterings, and of processes involving two and three body
initial or final states.  In Section 2 we derive the Boltzmann Equations (BE)
involving these terms and obtain a general parameterization for them.  In
Section 3 we present a detailed computation of the CP violating asymmetries in
scatterings in zero temperature field theory.  These asymmetries have been
considered before in the limit in which they are proportional to the CP
asymmetry in decays \cite{pi04-pi05,aba06b}, and we discuss here the validity
of this approximation. In Section 4 we stress how in the thermal
leptogenesis scenario, when one starts with a vanishing initial abundance for
the right handed neutrinos and assumes that no other interactions besides the
Yukawa couplings can produce them, washout processes are essential to generate
a lepton asymmetry. This is due to the fact that the lepton asymmetry produced
at early times (mainly through scatterings mediated by off-shell $N_1$ and
processes producing real $N_1$'s) tends to be compensated by the opposite sign
asymmetry produced at late times (when $N_1$ disappearance processes become
the dominant source).  The inclusion in the BE of CP violating asymmetries in
scatterings is important to make  this cancellation complete.  In the presence
of washouts, the cancellation can be partially avoided because washout
processes erase slightly more efficiently the asymmetry produced at early
times than the asymmetry produced at later times.  However, when washouts are
particularly weak the cancellation remains effective, and in this case the
inclusion of CP violation in scatterings leads to a sizeable suppression of
the final lepton asymmetry.  In the strong washout regimes washout processes
attain thermal equilibrium, thus erasing any dependence from the earlier
conditions and in particular from the large asymmetries generated by the
scattering processes. Since at lower temperatures the contributions from
scatterings are highly suppressed and the lepton asymmetry is essentially
generated through decays, for the strong washout 
 regimes the inclusion of scattering CP
asymmetries leaves the final results approximately unchanged.

\section{The Boltzmann equations}
\label{sec:BE}

We consider thermal leptogenesis scenarios with hierarchical heavy neutrinos
$N_i$, i.e. with masses $M_{2,3}\gg M_1$. We will focus on the evolution of
the lightest one, that in this section will be denoted simply by $N$, and we
will ignore the possible contributions from the heavier neutrinos
$N_{2,3}$~\cite{eng06} except for their virtual effects in the CP violating
asymmetries.

Let us first introduce some notation.  We denote the thermally averaged rate
for an initial state $A$ to go into the final state $B$ (summed over initial
and final spin and gauge degrees of freedom) as:
\begin{equation}
\label{eq:AtoB}
\gamma^A_B\equiv \gamma(A\to B), 
\end{equation}
and the CP difference between the processes for particles and  antiparticles as
\begin{equation}
\label{eq:DeltaAtoB}
\Delta \gamma^A_B\equiv \gamma^A_B-\gamma^{\bar A}_{\bar B}. 
\end{equation}
Particle densities are written in terms of the entropy density $s$, i.e.
$Y_a\equiv n_a/s$ where $n_a$ is the number density for the particle $a$. To
simplify the expressions we rescale the densities $Y_a$ by the equilibrium
density $Y_a^{eq}$ of the corresponding particle, defining $y_a\equiv
Y_a/Y_a^{eq}$, while the asymmetries of the rescaled densities are
denoted\footnote{The notation adopted here differs 
   from the notation used in \cite{na05,na06} in which the symbol $y_a$, rather
  than $\Delta y_a$, was used to denote the asymmetries of the rescaled
  densities.} by
$\Delta y_a\equiv y_a-y_{\bar a}$.

The difference between a process and its time reversed, weighted by 
the   densities of the initial state particles, is denoted as  
\begin{equation}
[A\leftrightarrow B]\equiv (\prod_{i=1}^{n}y_{a_i})\gamma^A_B-(\prod_{j=1}^{m}y_{b_i})
\gamma_A^B,
\end{equation}
where the state $A$ contains the particles $a_1, ... ,a_n$ while the state $B$
contains the particles $b_1, ..., b_m$.  We will consider only processes in
which at most one intermediate state heavy neutrino $N$ can go on the mass
shell, and in these cases a primed notation $\gamma{}'^A_B$ will refer to the
rates with the resonant intermediate state (RIS) subtracted (and similarly for
$[A\leftrightarrow B]'$).  Accordingly, a process $A \to B$ of this kind will
be divided into a RIS subtracted (off-shell) piece, and a second part
corresponding to on-shell $N$ exchange:
\begin{equation}
 \label{eq:gammap}
\gamma^A_B =\gamma'{}^A_B + {\gamma^{os}}^A_B.
  \end{equation}
  In the simple case when only $2\leftrightarrow 2$ scatterings are
  considered, the on-shell part is just
\begin{equation}
{\gamma^{os}}^A_B\equiv \gamma^A_N B^N_B,
\label{gamos}
\end{equation} 
where $B^N_B$ denotes the branching ratio for $N$ decays into the final state
$B$. (As is discussed in section~\ref{sec:1to3-2to3}, the inclusion of
$2\leftrightarrow 3$ scatterings, and in general of processes of higher order
in the couplings, implies that eq.~(\ref{gamos}) needs to be generalized.) 

In the following two sections we will write down the BE for the evolution of
the density of the heavy Majorana neutrinos $N$ and of the asymmetry for a
generic lepton flavor $i$.  We will first consider in
section~\ref{sec:1to2-2to2} the leading terms involving the Yukawa couplings
(generically denoted as $\lambda$) of the neutrino $N$ to the Higgs boson $H$
and to a light lepton doublet $\ell_i$.  In section~\ref{sec:1to3-2to3} we
will include the additional contributions arising from processes involving
also the top Yukawa coupling $h_t$ and the gauge interactions.  Accordingly,
the contributions to the evolution equation for the $N$ density will be
split in two parts:
\begin{equation}
  \label{eq:YN}
\dot Y_N =\left(\dot Y_N\right)_I+\left(\dot Y_N\right)_{II},   
\end{equation}
where we have introduced the  notation $\dot Y \equiv szH{\rm d}Y/{\rm d}z$,
with $z\equiv M_N/T$  and $H(z)$ being the Hubble rate  at temperature $T$. In
eq.~(\ref{eq:YN})  $\left(\dot  Y_N\right)_I$  includes the  contributions  of
terms up to ${\cal O}(\lambda^2)$, while $\left(\dot Y_N\right)_{II}$ includes
the  contributions up to  ${\cal O}(\lambda^2  h_t^2)$ or  ${\cal O}(\lambda^2
g^2)$, with $g$ a generic gauge coupling constant.

The contributions to the evolution equation for the density of the lepton
flavor $i$ will also be split in different parts: 
\begin{equation}
  \label{eq:YLi}
\dot Y_{L_i} =\left(\dot Y_{L_i}\right)_I+
\left(\dot Y_{L_i}\right)_{II}  +
\left(\dot Y_{L_i}\right)_{sphal},   
\end{equation}
with $Y_{L_i}\equiv 2 Y_{\ell_i}+Y_{e_i}$, where the factor 2 comes from
summing over the densities of the two gauge degrees of freedom in $\ell_i$,
and the inclusion of the density $Y_{e_i}$ for the right-handed lepton $e_i$
is required because the $L$-conserving charged lepton Yukawas can transfer
part of the asymmetry to the right handed degrees of freedom.  In
eq.~(\ref{eq:YLi}) $\left(\dot Y_{L_i}\right)_I $ includes the contributions
of terms up to ${\cal O}(\lambda^4)$, while $ \left(\dot Y_{L_i}\right)_{II} $
includes the contributions up to ${\cal O}(\lambda^4 h_t^2)$ or ${\cal
  O}(\lambda^4g^2)$.

The last term $\left(\dot Y_{L_i}\right)_{sphal}$ represents the change in the
lepton densities due to electroweak sphalerons, which are the only source of
baryon number violation.  Although the precise rates of sphaleron effects are
hard to estimate, one knows that they are efficient below $T\simeq
10^{12}\,$GeV and that they leave unchanged $B-L$.  Moreover, sphalerons
generate the same change in the baryon number of each generation, in fact one
has that $ (\dot{Y}_{\Delta L_i})_{sphal}=(\dot{Y}_{\Delta B})_{sphal}/3$,
where $ Y_{\Delta L_i}\equiv Y_{L_i}-Y_{\bar L_i}$ and $Y_{\Delta B}$ is the
baryon asymmetry to entropy ratio.  Hence, it is convenient to write directly
an equation for the quantities $Y_{\Delta_i}\equiv Y_{\Delta B}/3- Y_{\Delta
  L_i}$ which do not depend on the sphaleron rates.  By subtracting from
eq.~(\ref{eq:YLi}) the analogous equation for $\dot{Y} _{\bar L_i}$ and by
subtracting again the result from the equation that describes the evolution of
the baryon asymmetry, $(\dot{Y}_{\Delta B})/3= (\dot{Y}_{\Delta
  B})_{sphal}/3$, one obtains
\begin{equation}
  \label{eq:YDeltai}
\dot Y_{\Delta_i} =-\left(\dot Y_{\Delta L_i}\right)_I-
\left(\dot Y_{\Delta L_i}\right)_{II}\,.   
\end{equation}

 \subsection{$1\leftrightarrow 2$ and $2\leftrightarrow 2$ processes}
\label{sec:1to2-2to2} 

Let us briefly sketch the way in which the BE are obtained, by
considering  first the lepton (flavor) number violating $2\leftrightarrow 2$
scatterings mediated by $N$ exchange, that is
 $\ell_i H\leftrightarrow \ell_j H$ (with $j\neq i$) and  
$\ell_i H\leftrightarrow \bar\ell_j \bar H $.
  The BE describe the evolution of the particle densities at any
given time, but since the Universe is expanding, the temperature is decreasing
and the particle densities are changing, only processes that to a sufficiently
good approximation can be described by (effective) contact interactions can be
consistently included.  However, the $2\leftrightarrow 2$ scatterings we are
considering can proceed through the exchange in the $s$-channel of an on-shell
$N$, and in this case they are characterized by a time scale that can be
comparable to the expansion rate of the Universe (at $T\sim M_1$). Therefore
they cannot be approximated by a contact interaction, and must be treated with
care.  The usual way to deal with this is  to separate the off-shell part
of the scatterings from the on-shell piece, i.e.
\begin{equation}
\label{eq:2-2}
\gamma^{\ell_iH}_{\bar\ell_j \bar H }\equiv 
 \gamma'{}^{\ell_iH}_{\bar\ell_j \bar H } +
\gamma^{os}{}^{\ell_iH}_{\bar\ell_j \bar H }.
\end{equation}
and similarly for $\ell_i H\leftrightarrow \ell_j H$.  Then, at each given
time, the evolution of the density of the lepton flavor $L_i$ is determined by
the following (instantaneous) reactions:
\begin{enumerate}
\item[\it i)] $N \to \ell_i H$ decays occurring at a rate proportional to the
  density $Y_N$ of the right handed neutrinos, and $\ell_i H \to N$ inverse
  decays;
\item[\it ii)] off-shell $2\leftrightarrow 2$  scatterings
  $\gamma{}'^{\bar\ell_j\bar H}_{\ell_i H}$ and $\gamma{}'^{\ell_j H}_ {\ell_i
    H}$ involving only  the exchange of virtual $N$'s (in the  first process 
$N$ can be exchanged in both the $s$ and $t$ channels, while in the second process
  only in the $s$ channel);
\item[\it iii)] $2\leftrightarrow 2$ scatterings ${\gamma}^{\bar H \bar H}_
  {\ell_i \ell_j}$ and ${\gamma}^{ H \bar H}_ {\ell_i \bar \ell_j}$ with $N$
  exchanged in the $t$ and $u$-channels (in these cases no RIS can appear, and
  hence there are no on-shell contributions to be subtracted).
\end{enumerate}
The corresponding BE then reads: 
\begin{equation}
\nonumber \label{eq:YLi-I}
\left(\dot{Y} _{L_i}\right)_I =
\left(\dot{Y} _{L_i}\right)_{1\leftrightarrow 2} +
\left(\dot{Y} _{L_i}\right)_{2\leftrightarrow 2}^{\rm sub}+
\left(\dot{Y} _{L_i}\right)_{2\leftrightarrow 2}^{N_t}\,, 
\end{equation}
where  
\begin{eqnarray}
\label{eq:YLi1to2}
\left(\dot{Y} _{L_i}\right)_{1\leftrightarrow 2} &=& 
[N \leftrightarrow  \ell_i H],
\\  [4pt] 
\left(\dot{Y} _{L_i}\right)_{2\leftrightarrow  2}^{\rm sub} &=& 
\sum_j [\bar \ell_j \bar H \leftrightarrow \ell_i H ]'+
\sum_{j\neq i} [\ell_j H \leftrightarrow  \ell_i  H]' , \label{eq:YLi2to2sub}
\\ 
\left(\dot{Y} _{L_i}\right)_{2\leftrightarrow 2}^{N_t} &=& 
\sum_j \left\{
[H \bar H \leftrightarrow  \ell_i  \bar \ell_j]+
(1+\delta_{ij}) [\bar H \bar H \leftrightarrow \ell_i \ell_j ]
\right\} .
\label{eq:YLi2to2Nt}
\end{eqnarray}
It is important to remark that while eq.~(\ref{eq:YLi1to2}) contains processes
of ${\cal O}(\lambda^2)$, both eqs.~(\ref{eq:YLi2to2sub}) and
(\ref{eq:YLi2to2Nt}) contain only non-resonant scatterings, that are ${\cal
  O}(\lambda^4)$.  However, while in a first approximation the contributions
in eq.~(\ref{eq:YLi2to2Nt}) may be neglected, the inclusion of the off-shell
contributions of  eq.~(\ref{eq:YLi2to2sub}) is mandatory.  This is because the
CP asymmetries of the subtracted rates are of the same order than the CP
asymmetries of decays and inverse decays, and therefore neglecting them would
yield inconsistent results.  We can now subtract from eq.~(\ref{eq:YLi-I}) the
analogous equation for $Y_{\bar L_i}$ and write separately the source and
washout contributions to $Y_{\Delta L_i} \equiv Y_{L_i}-Y_{\bar L_i}$ as:
\begin{equation}
\label{eq:YDLi-I}
\left(\dot{Y}_{\Delta L_i}\right)_I = \left(\dot{Y}_{\Delta L_i}\right)^s_I+
\left(\dot{Y} _{\Delta L_i}\right)^w_I\,. 
\end{equation}
At the leading order, the source term receives contributions from the
$1\leftrightarrow 2$ decays and inverse decays in eq.~(\ref{eq:YLi1to2}) and
from the off-shell parts of the $2\leftrightarrow 2$ scatterings in
eq.~(\ref{eq:YLi2to2sub}), while the CP asymmetries of the $t$- and
$u$-channel processes in eq.~(\ref{eq:YLi2to2Nt}), being of higher order in
the couplings, can be neglected.  We can then write the source term as:

\begin{equation}
\label{eq:YLi-s-I}
\left(\dot{Y} _{\Delta L_i}\right)^s_I=
\left(\dot{Y} _{\Delta L_i}\right)^s_{1\leftrightarrow2}+
\left(\dot{Y} _{\Delta L_i}\right)^{s\>{\rm sub}}_{2\leftrightarrow2}, 
\end{equation}
where
\begin{eqnarray}
\label{eq:1to2s} 
&&\hspace{-1.3cm}\left(\dot{Y} _{\Delta L_i}\right)^s_{1\leftrightarrow2}=
(y_N+1)\Delta\gamma^N_{\ell_iH}, 
\\[4pt]
&&\hspace{-1.3cm}\left(\dot{Y}_{\Delta L_i}
\right)^{s\>{\rm sub}}_{2\leftrightarrow2}
= 
2\left(\sum_j\Delta\gamma'{}_{\ell_iH}^{\bar\ell_j\bar H }
+\sum_{j\ne i}\Delta\gamma'{}_{\ell_iH}^{\ell_jH}\right).
\label{eq:off-shell2-2}
\end{eqnarray}
In order to proceed we now use an important relation that states that the CP
asymmetries in the off-shell scatterings are, to leading order in the
couplings, equal in magnitude and opposite in sign with respect to the CP
asymmetries of the corresponding on-shell scatterings:
\begin{equation} \label{eq:on-off}
\Delta\gamma'{}^A_B\simeq -\Delta\gamma^{os}{}^A_B.
\end{equation}
To derive this relation, we first substitute the definition of the off-shell
rates in eq.~(\ref{eq:gammap}) to get $\Delta\gamma'{}^A_B\ =
\Delta\gamma^A_B- \Delta\gamma^{os}{}^A_B$ and then we use the fact that the
CP asymmetry of any process is always of higher order in the couplings with
respect to the corresponding tree level process \cite{Kolb:1979qa}. Thus the
CP asymmetries $\Delta \gamma^{\bar\ell_j \bar H }_{\ell_iH}$ and $\Delta
\gamma^{\ell_j H}_{\ell_i H }$ of the full $2\leftrightarrow 2$ scatterings
are ${\cal O}(\lambda^6)$. 
On the other hand, the CP asymmetries of the on-shell pieces are of ${\cal
  O}(\lambda^4)$.  This can be seen by writing them in terms of
eq.~(\ref{gamos}) to obtain
\begin{equation}\label{eq:Deltaos} 
\Delta\gamma^{os}{}^A_B=\Delta\gamma^A_N\  B^N_B+\gamma^A_N\frac{\Delta
  \gamma^N_B}{\gamma_{tot}}\simeq B_A^N\Delta \gamma^N_B-B^N_B\Delta\gamma^N_A.
\end{equation}
In eq.~(\ref{eq:Deltaos}) $\gamma_{tot}$ is the total $N$ decay rate, and we
have used $\Delta\gamma^A_N=\Delta\gamma_{\bar A}^N=-\Delta\gamma_A^N$ where
the first equality follows from the CPT relation $\gamma(A\to B)=\gamma(\bar
B\to \bar A)$, and we have approximated at leading order $B^N_{\bar A}\simeq
B^N_A$.  This shows that $\Delta\gamma^{os}{}^A_B$ (and hence
$\Delta\gamma'{}^A_B$) is of the same order in the couplings as the CP
asymmetry in decays $\Delta\gamma^N_A$ (that is ${\cal O}(\lambda^4)$).
Therefore, up to ${\cal O}(\lambda^6)$ corrections, the contributions of the
off-shell rates in eq.~(\ref{eq:off-shell2-2}) can be written as
\begin{eqnarray}
\left(\dot{Y}_{\Delta L_i}
\right)^{s\>{\rm sub}}_{2\leftrightarrow2}
 \simeq 
- 2\Delta\gamma^N_{\ell_i H}
\sum_j \left(B^N_{\ell_j H}+B^N_{\bar \ell_j\bar H}\right).
\label{eq:sub2to2s}
\end{eqnarray}
Here $B^N_{\ell_j H}$ 
represents the branching ratio for the decay $N\to \ell_j H$. 
At the order in the couplings we are working here $\sum_j \left(B^N_{\ell_j
    H}+B^N_{\bar \ell_j\bar H}\right)\simeq 1$ and therefore the r.h.s. of
eq.~(\ref{eq:sub2to2s}) can be further simplified to $-2\Delta
\gamma^N_{\ell_i H}$.
After summing up the two contributions (\ref{eq:1to2s}) and
(\ref{eq:sub2to2s}), the source term $\left(\dot{Y}_{L_i}\right)_I$ in
eq.~(\ref{eq:YLi-s-I}) becomes proportional to $y_N-1$.  This is in agreement
with the general condition that no asymmetry can be generated in thermal
equilibrium, and constitutes a check that, at this order, all the
relevant contributions to the source term of the BE have been included.

The washout term $\left(\dot{Y} _{\Delta L_i}\right)^w_I$ in
eq.~(\ref{eq:YDLi-I}) contains the terms proportional to the light particle
asymmetries, and is the sum of three different contributions 
obtained by subtracting from
eqs.~(\ref{eq:YLi1to2})-(\ref{eq:YLi2to2Nt})  the corresponding
equations for $Y_{\bar L_i}$. It can thus be written as
\begin{equation}
  \label{eq:eq:YLi-w-I}
\left(\dot{Y} _{\Delta L_i}\right)^w_I= 
 \left(\dot{Y}_{\Delta L_i}\right)_{1\leftrightarrow 2}^{w} +
  \left(\dot{Y}_{\Delta L_i}\right)_{2\leftrightarrow 2}^{w,\rm sub}+
\left(\dot{Y} _{\Delta L_i}\right)_{2\leftrightarrow 2}^{w,Nt}\,. 
\end{equation}
After linearizing in the CP asymmetries and in the asymmetries of the
normalized densities, these  contributions read:
\begin{eqnarray}
\label{eq:1to2w} 
&&\hspace{-1cm} \left(\dot{Y}_{\Delta L_i}\right)_{1\leftrightarrow 2}^{w} = 
- (\Delta y_{\ell_i}+\Delta y_H)  \gamma_N^{\ell_iH}, \\  \nonumber
&&\hspace{-1cm} 
\left(\dot{Y}_{\Delta L_i}\right)_{2\leftrightarrow  2}^{w,\rm sub} =
- \sum_j\left[ (\Delta y_{\ell_i}+\Delta y_H)
\left( \gamma'{}^{\ell_iH}_{\bar\ell_j\bar H }
+ {\gamma'}^{\ell_iH}_{\ell_jH}\right) \right.
  \\    [-8pt] 
\label{eq:sub2to2w}  
  &&\hspace{4cm} 
\left. +  
(\Delta y_{\ell_j}+\Delta y_H)\left( \gamma'{}^{\ell_iH}_{\bar\ell_j\bar H }-
\gamma'{}^{\ell_iH}_{\ell_jH}\right)\right], \\  \nonumber
&&\hspace{-1cm}  
\left(\dot{Y}_{\Delta L_i}\right)_{2\leftrightarrow 2}^{w,N_t} = - 
\sum_{j}\left[ (1+\delta_{ij})(\Delta y_{\ell_i}+\Delta y_{\ell_j}+2\Delta y_H)
\gamma^{\ell_i\ell_j}_{\bar H \bar H}
 \right. \\  [-12pt] 
 \label{eq:2to2Ntw}  
  &&\hspace{6cm} \left. 
+ (\Delta y_{\ell_i}-\Delta y_{\ell_j})
\gamma^{\ell_i\bar\ell_j}_{\bar H  H}
\right].
\end{eqnarray}
One can estimate the off-shell parts in
eq.~(\ref{eq:sub2to2w}) by introducing a subtracted propagator for $N$, or
alternatively eliminating the $\gamma'$ 
by means eq.~(\ref{eq:gammap}). 

Finally, to compute the lepton asymmetry we also need to solve for the
evolution of the heavy neutrino density $y_N$ that appears in the source term
eq.~(\ref{eq:1to2s}).  To the leading order in the couplings, the corresponding
BE reads
\begin{equation}\label{eq:YN-I}
\left(\dot{Y}_N\right)_I= \sum_j\Big\{[\ell_j H \leftrightarrow N] + 
[\bar \ell_j \bar H  \leftrightarrow N]\Big\}  
\simeq -(y_N-1)\gamma_D^{N\to 2}, 
\end{equation}
where $\gamma_D^{N\to 2} = \sum_j(\gamma^N_{\ell_j H} + \gamma^N_{\bar \ell_j
  \bar H})$ is the thermally averaged two body $N$ decay rate, and we have
approximated the leptons and Higgs particle densities with their equilibrium
values.

\vspace{.5cm} 

\subsection{$1\leftrightarrow 3$ and $2\leftrightarrow 3$ processes}
\label{sec:1to3-2to3}

We are now ready to generalize the above procedure to include processes
involving the Higgs Yukawa coupling $h_t$ to the right handed top quark, 
which is sizeable.  Processes involving gauge bosons can be included in an
entirely similar way and will be mentioned later.

The inclusion of $\Delta L=1$ processes like $N\leftrightarrow \ell_i \bar Q
t$ decays and inverse decays (where $Q$ is the left-handed quark doublet and
$t$ the right-handed top singlet) and scatterings mediated by Higgs exchange
like $N \ell_i \leftrightarrow Q \bar t$, follows along lines analogous to those
presented  in the previous section.  For the evolution of the 
heavy neutrino density we obtain  
\begin{equation}
\left(\dot{Y}_N\right)_{II}=
-(y_N-1)\left[\gamma_D^{N\to 3}+\gamma^{2\leftrightarrow 2}_{top}
\right],
\label{ypn}
\end{equation}
where
\begin{equation}
\gamma_D^{N\to 3}\equiv \sum_j(\gamma^N_{\ell_j\bar Q t}+\gamma^N_{\bar \ell_j
Q\bar t})
\end{equation}
is the contribution from  decays into three body final states, and  the
contribution from Higgs mediated scatterings is
\begin{equation}
\gamma^{2\leftrightarrow 2}_{top} = 
\sum_j\left( \gamma^{N\ell_j}_{Q\bar t}+
\gamma^{N\bar\ell_j}_{\bar Q t} + 
\gamma^{NQ}_{\ell_j t}+
\gamma^{N\bar Q}_{\bar\ell_j\bar t}+\gamma^{Nt}_{\bar\ell_jQ}+
\gamma^{N\bar t}_{\ell_j \bar Q}
\right), 
\end{equation}
where the first two terms in the sum correspond to the contributions from
$s$-channel Higgs exchange, while the other four terms (that at leading order
are all equal) correspond to the contribution from $t$ and $u$ channel Higgs
exchange.

Regarding the evolution of the lepton asymmetries, the derivation of the BE is
more delicate because, besides the inclusion of the CP violating asymmetries
in $1 \leftrightarrow 3$ decays like $N\leftrightarrow \ell_i\bar Q t$ and in
$2\leftrightarrow 2$ scatterings like $N \ell_i \leftrightarrow Q \bar t $,
the asymmetries of various off-shell $2\leftrightarrow 3$ scatterings,
that contribute to the source term at the same order in the couplings, should
also be included. Accordingly, the term $\left(\dot Y_{L_i}\right)_{II}$ in
eq.~(\ref{eq:YLi}) can be written as
\begin{equation}
\left(\dot{Y}_{L_i}\right)_{II} = 
\left(\dot{Y}_{L_i}\right)_{\stackrel{\scriptstyle1 \leftrightarrow  3}{\scriptstyle 2 \leftrightarrow  2}}
+\left(\dot{Y}_{L_i}\right)_{2\leftrightarrow 3}^{\rm sub} + 
\left(\dot{Y}_{L_i}\right)_{2\leftrightarrow 3}^{N_t}
\end{equation}
with  
\begin{eqnarray}
\label{eq:1to3}
\left(\dot{Y}_{L_i}\right)_{\stackrel{\scriptstyle1 \leftrightarrow
    3}{\scriptstyle 2 \leftrightarrow  2}}
\!\!\! &=& \!\!\!
[N\leftrightarrow {\ell_i \bar Q t}]+ [{Q\bar  t}\leftrightarrow {N\ell_i}]+  
[N\bar t\leftrightarrow \bar Q\ell_i]+[N Q\leftrightarrow  t\ell_i]\,; 
\\[6pt]\nonumber
\left(\dot{Y}_{L_i}\right)_{2\leftrightarrow 3}^{\rm sub} 
\!\!\!&=&\!\!\!
\sum_{j\neq i}\left\{
[\ell_j H\leftrightarrow \bar Q t\ell_i]'+
[\ell_j HQ\leftrightarrow  t\ell_i]'+
[\ell_j H\bar t\leftrightarrow \bar Q \ell_i]' 
\right. 
\\[-8pt]\nonumber   
&& \hspace{1.cm}
+\left.
 [\ell_j \bar Q t \leftrightarrow  \ell_i H]' +  
[\ell_j \bar Q  \leftrightarrow  \ell_i H \bar t]' +
[\ell_j t \leftrightarrow  \ell_i H Q]' 
         \right\}  \qquad\\ [4pt] \nonumber
&& \hspace{-1.4cm} + \sum_j\left\{
[\bar \ell_j\bar H\leftrightarrow \ell_i \bar Qt]'+
[\bar\ell_j \bar H Q\leftrightarrow  t\ell_i]'+
[\bar \ell_j \bar H\bar t\leftrightarrow \bar Q \ell_i]'+
  [\bar\ell_jQ \bar t\leftrightarrow \ell_i H]'
\right. 
\\[-2pt]\nonumber   
 && \hspace{-1. cm} 
+ \left.
[\bar\ell_jQ \leftrightarrow \ell_i Ht]'+
[\bar\ell_j \bar t\leftrightarrow \ell_i H\bar Q]'+
[\bar Q  t\leftrightarrow \ell_i H\bar \ell_j]'+
[Q \bar t\leftrightarrow \ell_i \bar H\bar \ell_j]'
 \right\} 
 \\[6pt]  
\label{eq:Lsub2to3}
 && \hspace{-1.2cm} 
+ \sum_j(1+\delta_{ij})         
[Q \bar t\leftrightarrow \ell_i H\ell_j]'\,;
\\[2pt]  \nonumber
\left(\dot{Y}_{L_i}\right)_{2\leftrightarrow 3}^{N_t} 
\!\!\!\!&=& \!\!\!\!
 \sum_{j\neq i}\left\{ 
[\ell_j Q \bar t\leftrightarrow \ell_i \bar H] +
[\ell_j\bar H\leftrightarrow \ell_i Q\bar t] +
[\ell_j\bar H t\leftrightarrow \ell_i Q]+
[\ell_j\bar H\bar Q\leftrightarrow \ell_i \bar t] \right\}
 \\[-2pt]  \nonumber
&& \hspace{-1.3cm} +\sum_{j}\left\{ 
[Q\bar t  H\leftrightarrow \bar \ell_j \ell_i] +
[\bar Q t \bar H\leftrightarrow \bar \ell_j \ell_i] +
[\bar Q \bar H\leftrightarrow \bar \ell_j \ell_i\bar t ] +
[t \bar H\leftrightarrow \bar \ell_j \ell_i\bar Q ]
\right\} 
\\[-2pt] 
&&\hspace{-1.3cm} + \sum_j  (1+\delta_{ij})\left\{
[\bar t \bar H \leftrightarrow \ell_i \bar Q\ell_j]+
[Q \bar t\bar H\leftrightarrow \ell_i \ell_j]+
[Q \bar H\leftrightarrow \ell_i t\ell_j]
\right\}\,.
\label{eq:2to3Nt}
\end{eqnarray}
As in the previous section, the asymmetries of the off-shell $2\leftrightarrow
3$ rates in eq.~(\ref{eq:Lsub2to3}) can be estimated by relating them to the
asymmetries of the corresponding on-shell parts.  However, for
$2\leftrightarrow 3$ scatterings the definition of the on-shell part is more
subtle, because after a real $N$ is produced in a collision, it has a certain
probability to scatter before decaying.  Let us consider for example the
process ${\ell_j H Q}\to {t \ell_i}$.  The contribution to this process
from the exchange of an on-shell $N$ corresponds to the production process
$\ell_j H \to N$ followed by the scattering $N + Q \to t  \ell_i$ mediated
by a Higgs in the $t$ channel.  Processes of this kind can generally be
written as $AX\to Y$ where $A$ denotes a possible state to which $N$ can decay.
The corresponding on-shell rate then is 
\begin{equation}\label{eq:os2b}
\gamma^{os}{}^{AX}_Y = \gamma^A_NP^{NX}_Y,\\ 
\end{equation}
where we have introduced the quantity $P^{NX}_Y$ that is the probability that
the heavy neutrino $N$ will scatter with $X$ to produce $Y$. Processes in
which the on-shell $N$ can disappear only by decaying (as for example $\ell_j
H \leftrightarrow \ell_i \bar Q t$ or $\ell_j t \leftrightarrow \ell_i H Q$)
can generally be written as $A \to B$ or as $X\to B Y$ where both $A$ and $B$
denote possible final states for $N$ decays.  The corresponding on-shell rates
are
\begin{eqnarray}
\label{eq:os2a}
\gamma^{os}{}^A_B &=& \gamma^A_NP^N_B,\\ 
\label{eq:os2c}
\gamma^{os}{}^{X}_{BY} &=& \gamma^X_{NY}P^N_B. 
\end{eqnarray}
Note that because of the fact that in the dense plasma $N$ can suffer
inelastic scatterings before decaying as is described by eq.~(\ref{eq:os2b}), the
quantities $P^N_B$ in eqs.~(\ref{eq:os2a}) and (\ref{eq:os2c}) differ from the
usual notion of branching ratios at zero temperature.  In particular,
scattering rates should also be included in normalizing properly the decay
probabilities, and the quantities $P^a_b$ then will denote the general
probabilities that the heavy neutrino $N$ contained in state $a$ ends up
producing state $b$.  In the case under discussion, we have for example
\begin{eqnarray} 
 \nonumber
 && 
P^N_{\ell_i H}=\frac{\gamma^N_{\ell_i H}}{\gamma_{all}}, 
\quad\ 
P^N_{\ell_i \bar Q t}=\frac{\gamma^N_{\ell_i \bar Q t}}{\gamma_{all}}, 
 \\ \label{eq:PN}
 && 
P^{N\ell_i}_{Q\bar  t}=
 \frac{ \gamma^{N\ell_i}_{Q\bar  t}}{\gamma_{all}},  \quad
P^{NQ}_{t \ell_i}= 
\frac{ \gamma^{N Q}_{t\ell_i}}{\gamma_{all}},  \quad
P^{N\bar t}_{\bar Q \ell_i}=
 \frac{\gamma^{N \bar t}_{\bar Q\ell_i}}{\gamma_{all}},
\end{eqnarray}
with similar definitions for the probabilities of the CP conjugate processes.
The probabilities are  normalized in terms of the sum  of 
all the rates,  that reads
\begin{equation} \label{eq:PNtot}
\gamma_{all}=\sum_i(\gamma^N_{\ell_i H}+\gamma^N_{\bar\ell_i \bar H}
+\gamma^N_{\ell_i \bar Qt}+\gamma^N_{\bar\ell_i Q \bar t}+
\gamma^{N\ell_i}_{Q \bar t}+\gamma^{N\bar\ell_i}_{\bar Q t}+
\gamma^{N\bar Q}_{\bar \ell_i\bar t}+\gamma^{N Q}_{\ell_i t}+
\gamma^{N\bar t}_{\ell_i\bar Q}+\gamma^{N t}_{\bar \ell_i  Q}).
\end{equation}
To the order in the Yukawa couplings that we are considering, the unitarity
condition for the sum of the branching ratios of $N$ into all possible final
states $\sum_Y B^N_Y=1$ is then generalized to $\sum_{X,Y} P^{NX}_Y=1$. That is,
the probabilities for all the possible ways through which $N$ can disappear
add up to unity.

To include the new sources of CP asymmetries, we now need to subtract from
eqs.~(\ref{eq:1to3})-(\ref{eq:2to3Nt}) the analogous equation for $Y_{\bar
  L_i}$. We obtain
\begin{equation}
\label{eq:YDLi-II}
(\dot{Y}_{\Delta L_i})^s_{II}= 
(\dot{Y}_{\Delta L_i})^s_
{\stackrel{\scriptstyle1 \leftrightarrow  3}{\scriptstyle 2 \leftrightarrow  2}}
+\left(\dot{Y}_{\Delta L_i}\right)_{2\leftrightarrow 3}^{s,\>\rm sub},  
\end{equation}
where we have neglected the CP asymmetries of the $2\leftrightarrow 3$
processes with $N$ exchanged in the $t$-channel  since
they are of higher order in the couplings.  For the first term in the r.h.s.
of eq.~(\ref{eq:YDLi-II}) we have
\begin{equation}
\label{eq:2to2s}
(\dot{Y} _{\Delta L_i})^s_{\stackrel{\scriptstyle1 \leftrightarrow  3}
{\scriptstyle 2 \leftrightarrow
    2}} = 
(y_N+1)\left[
\Delta\gamma^N_{\ell_i\bar Q t}+\Delta\gamma^{N\bar t}_{\ell_i\bar
  Q}+\Delta\gamma^{NQ}_{\ell_it} -\Delta\gamma^{N\ell_i}_{ Q\bar t}
\right].
\end{equation}
After eliminating the subtracted rates by writing their CP asymmetries as
minus the CP asymmetries of the on-shell rates and keeping terms up to ${\cal
  O}(\lambda^4h_t^2)$, we obtain for the second term in eq.~(\ref{eq:YDLi-II})
\begin{eqnarray}
\nonumber 
\left(\dot{Y}_{\Delta L_i}\right)_{2\leftrightarrow 3}^{s,\>\rm sub}\! &=& \! 
- 2\, \Delta\gamma^N_{\ell_i H}
\left[ 1 - \sum_j \left(P^N_{\ell_j H}+P^N_{\bar \ell_j\bar H}\right)\right]
  \\ \label{eq:sub2to3s}  
&&  \hspace{.8cm}
- 2 \left[
\Delta\gamma^N_{\ell_i\bar Q t}+\Delta\gamma^{N\bar t}_{\ell_i\bar
  Q}+\Delta\gamma^{NQ}_{\ell_it} -\Delta\gamma^{N\ell_i}_{ Q\bar t}
\right]
\,.
\end{eqnarray}
Note that at order ${\cal O}(\lambda^4 h_t^2)$  eq.~(\ref{eq:sub2to2s})  reads 
\begin{equation}
\label{eq:sub2to2sb}
\left(\dot{Y} _{\Delta L_i}\right)^{s\>{\rm sub}}_{2\leftrightarrow2} 
= - 2\Delta\gamma^N_{\ell_i H}
\sum_j \left(P^N_{\ell_j H}+P^N_{\bar \ell_j\bar H}\right),        
\end{equation}
where the sum of the probabilities $\sum_j \left(P^N_{\ell_j H}+P^N_{\bar
    \ell_j\bar H}\right)$ is not unity. However, the first term in
eq.~(\ref{eq:sub2to3s}) conspires with eq.~(\ref{eq:sub2to2sb}) to yield the
correct behavior for the source term involving $\Delta \gamma^N_{\ell_i H}$,
that turns out again to be proportional to $(y_N-1)$.  By summing up
eqs.~(\ref{eq:1to2s}), (\ref{eq:2to2s}), (\ref{eq:sub2to3s}) and
(\ref{eq:sub2to2sb}) we obtain the final expression for the source term that
holds at ${\cal O}(\lambda^4 h_t^2)$:
\begin{equation}
(\dot Y_{\Delta L_i})^s_{I+II}=(y_N-1)\left[\Delta\gamma^N_{\ell_iH}+
\Delta\gamma^N_{\ell_i\bar Q t}+\Delta\gamma^{N\bar t}_{\ell_i\bar
  Q}+\Delta\gamma^{NQ}_{\ell_it} -\Delta\gamma^{N\ell_i}_{ Q\bar t}
\right]. 
\end{equation}

Regarding the washouts, the contributions from 
eqs.~(\ref{eq:1to3})-(\ref{eq:2to3Nt})  after subtracting
 the analogous equations for 
$\dot Y_{\bar L_i}$ can be written as  
\begin{equation}
  \label{eq:eq:YLi-w-II}
\left(\dot{Y} _{\Delta L_i}\right)^w_{II}= 
 \left(\dot{Y}_{\Delta L_i}\right)^w_{\stackrel
{\scriptstyle1 \leftrightarrow 3}{\scriptstyle 2 \leftrightarrow  2}}+
\left(\dot{Y}_{\Delta L_i}\right)_{2\leftrightarrow 3}^{w,\> \rm sub} +
\left(\dot{Y}_{\Delta L_i}\right)_{2\leftrightarrow 3}^{w,\>N_t},  
\end{equation}
where 
\begin{eqnarray}  
 \nonumber
&& \hspace{-2cm} 
\left(\dot{Y}_{\Delta L_i}\right)^w_{\stackrel
{\scriptstyle1 \leftrightarrow 3}{\scriptstyle 2 \leftrightarrow  2}}
 =
\left[(\Delta y_Q-\Delta y_t-\Delta  y_{\ell_i})\gamma^N_{\ell_i \bar Q t}+
(\Delta y_Q-\Delta y_t-y_N\Delta y_{\ell_i})\gamma^{Q\bar t}_{N\ell_i}
\right.
 \\ [5pt]  
\label{eq:1to3w}
&& \hspace{-.0cm}+\left. 
(\Delta y_Q-y_N \Delta y_t-\Delta y_{\ell_i})\gamma^{N\bar t}_{\bar Q \ell_i}
+(y_N\Delta y_Q-\Delta y_t-\Delta y_{\ell_i})\gamma^{N Q}_{t \ell_i}
\right]\,;
 \\ [5pt] 
\nonumber
&& \hspace{-2cm} 
\left(\dot{Y}_{\Delta L_i}\right)_{2\leftrightarrow 3}^{w,\> \rm sub} 
=
\sum_{j\neq i}
\left[
 (\Delta y_{\ell_j}
 +\Delta y_Q
 -\Delta y_t
+\Delta y_H
-\Delta y_{\ell_i})
\left(
{\gamma{}'^{\ell_j H}_{\bar Q t\ell_i  }}\!\!+\!
{\gamma{}'^{\ell_j H Q }_{t \ell_i }}\!\!+\!
{\gamma{}'^{\ell_j H \bar t}_{\bar Q\ell_i}} 
\right) 
\right.
\\  \nonumber
&& \hspace{-.0cm}
+\left.
(\Delta y_{\ell_j}
-\Delta y_Q
+\Delta y_t
-\Delta y_H
-\Delta y_{\ell_i})
\left(
{\gamma{}'^{\ell_j \bar Q t}_{H \ell_i  }}+
{\gamma{}'^{\ell_j \bar Q }_{H\bar t \ell_i }}+
{\gamma{}'^{\ell_j  t}_{Q H \ell_i}}
\right) \right]
\\ [4pt]  \nonumber
&&\hspace{-1.cm}
+\sum_{j}
\left[
(-\Delta y_{\ell_j}
+\Delta y_Q
-\Delta y_t
-\Delta y_H
-\Delta y_{\ell_i})\times 
\right.
\\   \nonumber
&& \hspace{-.5cm}
\left.
\left(
{\gamma{}'^{\bar\ell_j \bar H}_{\bar Q t\ell_i  }}+
{\gamma{}'^{\bar \ell_j \bar H Q }_{t \ell_i }}+
{\gamma{}'^{\bar \ell_j \bar H \bar t}_{\bar Q\ell_i}}+
{\gamma{}'^{\bar \ell_j Q \bar t }_{H \ell_i }}+
{\gamma{}'^{\bar \ell_j Q  }_{t H \ell_i }}+
{\gamma{}'^{\bar \ell_j  \bar t }_{\bar Q H \ell_i }\!\!}
+ (1+\delta_{ij})
{\gamma{}'^{Q\bar  t }_{\ell_j H \ell_i }\!}
\right) 
\right.
\\ [3pt]  \nonumber
&&\hspace{-.0cm}
+\left.
(\Delta y_{\ell_j}
-\Delta y_Q
+\Delta y_t
-\Delta y_H
-\Delta y_{\ell_i})
{\gamma{}'^{\bar Q  t}_{\bar\ell_j H\ell_i}}
\right.
\\ [3pt] \label{eq:sub2to3w}
&&\hspace{2.cm}
+\left.
 (\Delta y_{\ell_j}
 +\Delta y_Q
 -\Delta y_t
+\Delta y_H
-\Delta y_{\ell_i})
{\gamma{}'^{Q \bar t}_{\bar\ell_j\bar H\ell_i}}\>
\right];
\\ [8pt]  
\nonumber
&& \hspace{-2cm} 
\left(\dot{Y}_{\Delta L_i}\right)_{2\leftrightarrow 3}^{w,\>N_t} 
=
\sum_{j\neq i}
\left[
(
\Delta y_{\ell_j}
+\Delta y_Q
-\Delta y_t
+\Delta y_H
-\Delta y_{\ell_i})
{\gamma^{\ell_j Q \bar t }_{ \bar H \ell_i }}
\right.
\\ [0pt] \nonumber
&& \hspace{-.0cm} + \left. \right(
\Delta y_{\ell_j}
-\Delta y_Q
+\Delta y_t
-\Delta y_H
-\Delta y_{\ell_i})
({\gamma^{\ell_j \bar H }_{ Q \bar t \ell_i }}+
{\gamma^{\ell_j \bar H t }_{ Q  \ell_i }}+
{\gamma^{\ell_j \bar H \bar Q}_{ \bar t \ell_i }}\left)
\Big]\right.  \\ \nonumber
&& \hspace{-1.5cm}
+ \sum_j
\left[
(
\Delta y_{\ell_j}
+\Delta y_Q
-\Delta y_t
+\Delta y_H
-\Delta y_{\ell_i})
{\gamma^{Q \bar t H }_{ \bar \ell_j \ell_i }}
 \right.
 \\ [-4pt]\nonumber
 && \hspace{-.5cm}
 \left. 
+
(
\Delta y_{\ell_j}
+\Delta y_Q
+\Delta y_t
-\Delta y_H
-\Delta y_{\ell_i})
{\gamma^{t \bar H }_{ \bar \ell_j \bar Q \ell_i }}
 \right.
 \\ [2pt] \nonumber
 &&  
+ \left. (
\Delta y_{\ell_j}
-\Delta y_Q
+\Delta y_t
-\Delta y_H
-\Delta y_{\ell_i})
\left(
{\gamma^{\bar Q t \bar H }_{ \bar \ell_j  \ell_i }}
+ {\gamma^{\bar Q  \bar H }_{ \bar \ell_j \bar t \ell_i }}
\right)
\right]
\\ [2pt] 
\label{eq:2to3Ntw}
&& \hspace{-2.2cm} 
+ \sum_j\left\{ 
(1+\delta_{ij}) 
(
-\Delta y_{\ell_j}
+\Delta y_Q
-\Delta y_t
-\Delta y_H
-\Delta y_{\ell_i})
\left(
{\gamma^{\bar t\bar H }_{ \ell_j \bar Q   \ell_j}}+
{\gamma^{Q \bar t\bar H }_{ \ell_j    \ell_j}}+
{\gamma^{Q \bar H }_{ \ell_j  t  \ell_j}}\right) 
\right\}.
\end{eqnarray}

We can now summarize (and slightly generalize) the procedure for writing the BE
for leptogenesis.  Ignoring processes with more than one $N$, one can write
them as 
\begin{eqnarray}
\label{eq:yNclosed}  
 &&\hspace{-.5cm} 
\dot{Y}_N =  -(y_N-1)\sum_{A,B} \gamma^{NA}_B, \\ 
 &&\hspace{-1.3cm} 
(\dot{Y}_{\Delta L_i})^s  = 
(y_N-1)\sum_{A,B}\left[L_i(B)-L_i(A)\right]\Delta\gamma^{NA}_B , 
\label{eq:yLisclosed} 
\end{eqnarray}
\begin{equation}
(\dot{Y}_{\Delta L_i})^w= \sum_{A,B}\left[L_i(B)-L_i(A)\right]
\left(y_N^{n_A}\sum_{a_i}\Delta y_{a_i}
-y_N^{n_B}\sum_{b_i}\Delta y_{b_i}\right)\gamma'{}^A_B,
\label{eq:ypwashht}
\end{equation}
where in the first two equations the states $A$ and $B$ contain only light
standard model particles, while in the last one
  for the case in which no on-shell intermediate $N$ is
allowed in the process $A\to B$, one has simply $\gamma'{}^A_B=\gamma{}^A_B$.
In these equations $L_i(A)$ denote the $i$ lepton flavor number of state $A$
while $n_A=0$ or 1 counts the number of $N$'s contained in state $A$ (and
similarly for $L_i(B)$ and $n_B$).  To avoid double counting in both
equations the sums are restricted to $L_i(B)> 0$ and $L_i(B)> L_i(A)$, where
the first restriction avoids double counting the CP conjugate processes with
$\bar \ell_i$ in the final states, and the second restriction avoids double
counting the time reversed processes where the number of $\ell_i$ in the
initial state is larger than in the final state.  In the equation
(\ref{eq:ypwashht}) for the washouts the $a_i$'s and $b_i$'s denote all the
particle species with non-vanishing asymmetries
contained respectively in states $A$ and $B$.  
  Let us note that in the cases
we have been discussing all the CP violating asymmetries appearing in
eq.~(\ref{eq:yLisclosed}) have $L_i(B)-L_i(A)=1$, and thus the expression for
the source term can be accordingly simplified.  However, this is not true for
the washout term since some  processes with
$L_i(B)-L_i(A)=2$ also contribute.

Several contributions that we have included for consistency and for
completeness, for practical purposes can be neglected without affecting
sizably the numerical results. This is the case for example for all the
non-resonant $2\leftrightarrow 3$ washout terms.  Also the inclusion of $N$
decays into three body final states has been carried out mainly for the sake
of completeness, that is to account for all the processes of the same order in
the couplings, and also to incorporate consistently those $2\leftrightarrow 3$
scatterings for which the on-shell piece involves precisely a $1 \to 3$ decay,
as for example $\ell_j H \to N \to \ell_i \bar Q t$.  However, the overall
impact on quantitative results of the three-body decay CP asymmetry
$\Delta\gamma^N_{\ell_i \bar Q t}$ (as well as the contribution to the
washouts of $3\to 1$ inverse decays) are rather small. This is because while
e.g.  $\Delta\gamma^N_{\ell_i \bar Q t}$ involves the same CP violating phase
as $\Delta\gamma^N_{\ell_i H}$, it also has a significant suppression factor
arising from three body phase space. 
The explicit expression for the three body decay rate is presented in the
Appendix. 

Following the same procedure outlined above, it is possible to include in the
BE other relevant processes, such as those involving the gauge
bosons~\cite{pi04-pi05,gi04}.  With all the subdominant terms neglected and
with the effects of the gauge bosons included, the simplified expression of
the BE for the evolution of $Y_N$ reads:
\begin{eqnarray}
  \label{eq:finalBEYN}
\dot Y_N =   
-(y_N-1)\left[ \gamma_D^{N\to 2 } + \gamma^{2\leftrightarrow 2}_{top} +
  \gamma^{2\leftrightarrow 2}_A
\right],
\end{eqnarray}
where  the term involving the gauge bosons is defined as
\begin{equation}
\label{eq:AsAtAu}
\gamma^{2\leftrightarrow 2}_A= 
\sum_j\left(\gamma^{N\ell_j}_{A\bar H}+\gamma^{N\bar\ell_j}_{A  H}
+\gamma^{N\bar H}_{A\ell_j}+\gamma^{N H}_{A  \bar \ell_j} + 
\gamma^{N A}_{\ell_j H}+\gamma^{N A}_{\bar \ell_j  \bar H}\right),
\end{equation}
where $A=W_i$ or $B$ for SU(2) and U(1) bosons respectively, and a sum over
all the gauge boson degrees of freedom is understood.  In
eq.~(\ref{eq:finalBEYN}) (as well as in eq.~(\ref{eq:finalBEYDeltai-s}) below)
we have neglected three-body decays like $\gamma^N_{AH\ell_i}$ involving the
gauge bosons because, similarly to the three-body decays involving the top
quarks, they are suppressed by phase space factors and give negligible
contributions, and we have also neglected the contributions to the washouts
from $2\leftrightarrow 3$ processes.  The simplified expression for the
evolution equation for the charge $Y_{\Delta_i} =Y_{\Delta B}/3-Y_{\Delta
  L_i}$ that is conserved by sphaleron interactions, is:
\begin{equation}
  \label{eq:finalBEYDeltai}
\dot Y_{\Delta_i}= 
 \left(\dot Y_{\Delta_i}\right)^s +  \left(\dot Y_{\Delta_i}\right)^w,
\end{equation}
with  
\begin{eqnarray}
&& \hspace{-1cm} \left(\dot Y_{\Delta_i}\right)^s =  - (y_N-1) 
 \left[\Delta\gamma^N_{\ell_iH}
+\Delta\gamma^{N\bar t}_{\ell_i\bar
  Q}+\Delta\gamma^{NQ}_{\ell_it}\right. \\
\nonumber
&& \hspace{2cm}\left.  -\Delta\gamma^{N\ell_i}_{ Q\bar t} -\Delta\gamma^{N\ell_i}_{A\bar H}
+\Delta\gamma^{N\bar H}_{A\ell_i}
+\Delta\gamma^{N A}_{\ell_i H}
\right]. 
 \label{eq:finalBEYDeltai-s}
\end{eqnarray}
\begin{eqnarray}
\nonumber
&& \hspace{-1cm} \left(\dot Y_{\Delta_i}\right)^w =
 (\Delta y_{\ell_i}+\Delta y_H)  \gamma_N^{\ell_iH}  \\  \nonumber
&&\hspace{0cm} 
+ \sum_j\left[ (\Delta y_{\ell_i}+\Delta y_H)
\left( \gamma'{}^{\ell_iH}_{\bar\ell_j\bar H }
+ {\gamma'}^{\ell_iH}_{\ell_jH}\right) 
 +  
(\Delta y_{\ell_j}+\Delta y_H)\left( \gamma'{}^{\ell_iH}_{\bar\ell_j\bar H }-
\gamma'{}^{\ell_iH}_{\ell_jH}\right)\right] \\ 
&&\hspace{0cm}  
+\sum_{j}\left[ (1+\delta_{ij})(\Delta y_{\ell_i}+\Delta y_{\ell_j}+
2\Delta y_H)
\gamma^{\ell_i\ell_j}_{\bar H \bar H}
 + (\Delta y_{\ell_i}-\Delta y_{\ell_j})
\gamma^{\ell_i\bar\ell_j}_{\bar H  H}\right]
\\\nonumber &&\hspace{0cm}
+(y_N\Delta y_{\ell_i}-\Delta y_Q+\Delta y_t)\gamma^{N \ell_i}_{Q\bar t}
-(y_N\Delta y_Q-\Delta y_t-\Delta y_{\ell_i}) \gamma^{N Q}_{t \ell_i}
\\\nonumber&&\hspace{0cm}
-(-y_N\Delta y_t+\Delta y_Q-\Delta y_{\ell_i}) \gamma^{N\bar t}_{\bar Q
  \ell_i}
-(y_N\Delta y_{\ell_i}+\Delta y_H)\gamma^{N \ell_i}_{A\bar H}
\\\nonumber&&\hspace{0cm}
+(y_N\Delta y_H+\Delta y_{\ell_i}) \gamma^{N H}_{A \bar \ell_i}
+(\Delta y_H+\Delta y_{\ell_i}) \gamma^{N A}_{H \ell_i}.
\end{eqnarray}

Before concluding this section one more remark is in order.  To solve the set
of coupled BE for the different lepton flavors it is necessary to express all
the normalized particle asymmetries $\Delta y_{a}$ and $\Delta y_H$ in terms
of the relevant charge asymmetries $Y_{\Delta_i}$ (and use the relation
$\Delta y_Q-\Delta y_t=\Delta y_H/2$).  In doing so one has to take into
account the constraints on chemical potentials enforced by the reactions that,
in the particular temperature range in which leptogenesis occurs, are faster
than the Universe expansion \cite{na05,ha90}.  Although we have written the
equations for a generic asymmetry $Y_{\Delta_i}$ of one of the three flavors,
it is important to keep in mind that when no lepton Yukawa couplings are in
equilibrium ($T>10^{12}$~GeV), one has just an effective one-flavor equation
for the asymmetry in the lepton $\ell$ coupled to the lightest heavy neutrino
$N$.  When just the $\tau$ Yukawa coupling is in equilibrium ($10^{9}\ {\rm
  GeV}<T<10^{12}$~GeV) two flavor components are relevant, $\ell_\tau$ and
$\ell_a$, where $\ell_a$ is the component of the lepton doublet produced in
$N$ decays which is orthogonal to $\ell_\tau$~\cite{na06}.  When also the
$\mu$ Yukawa coupling is in equilibrium ($T<10^{9}$~GeV) the three flavor
components of $\ell$ are completely projected out by the Yukawa interactions,
and the full set of equations for the three $Y_{\Delta_i}$, with
$i=e,\,\mu,\,\tau$, is needed.


\section{The CP asymmetry in scattering processes}
\label{sec:CPasymmetries}

We will now study the CP asymmetries in $2\rightarrow 2$ scattering processes
involving the top quark and with a Higgs exchanged in the $s$ or in the $t$
channels.  These asymmetries were considered previously in \cite{pi04-pi05}
and \cite{aba06b} where some arguments were given in support of an
approximate equality between the scatterings and the decay asymmetries, 
as for example:
 \begin{equation}
\frac{\Delta\gamma^{Q\bar t}_{N\ell_i}}{\gamma^{Q\bar t}_{N\ell_i}}\simeq 
\frac{\Delta\gamma^{N}_{\ell_i H}}{\gamma^{N}_{\ell_i H}}.
\end{equation}
In this section we will explore the validity of this kind of approximations.

The Lagrangian for the Yukawa interactions relevant for the computation of the
associated CP asymmetries, written in the mass eigenstate basis of the heavy
neutrinos $N_\alpha (\alpha = 1,2,3)$, of the charged leptons ($i=e,\mu,\tau$)
and of the quarks, reads:
\begin{equation}
\label{yuk}
{\cal L}_Y=
-\lambda_{i \alpha}\,\overline{N}_\alpha \ell_i\,  {\widetilde H}^\dagger - 
h_t\,\overline{Q}  t\,  {\widetilde H}+h.c.\ .
\end{equation}
Here $H=(H^+,H^0)^T$ is the Higgs field, with $\widetilde H =i\tau_2 H^*$.
Like in the usual case of the CP asymmetries in $N$ decays, the CP asymmetries
in scattering processes arise from the interference between the tree level
and one loop amplitudes. They  can be computed by explicit evaluation of the
corresponding loop integrals, or just by using the Cutkowski rules that give
directly the absorptive part of the Feynman diagrams.  

\begin{figure}[t!]
\begin{center}
\SetScale{1}
\begin{picture}(400,100)(-20,0)
\SetOffset(0,0)
%
\Text(45,0)[]{$(a)$}
\ArrowLine(-5,80)(20,50)
\ArrowLine(-5,20)(20,50)
\Text(5,85)[]{$Q$}
\Text(5,20)[]{$\bar t$}
\DashLine(20,50)(70,50){2}
\Text(45,60)[]{$H$}
\ArrowLine(70,50)(95,20)
\SetWidth{1.5}
\Line(70,50)(95,80)
\SetWidth{.5}
\Text(102,85)[]{$N_\alpha$}
\Text(102,20)[]{$\ell_i$}
\Text(180,0)[]{$(b)$}
\ArrowLine(130,80)(155,50)
\ArrowLine(130,20)(155,50)
\Text(140,85)[]{$Q$}
\Text(140,20)[]{$\bar t$}
\DashLine(155,50)(205,50){2}
\Text(180,60)[]{$H$}
\SetWidth{1.5}
\Line(205,50)(222,30)
\SetWidth{0.5}
\ArrowLine(222,30)(230,20)
\ArrowLine(205,50)(222,70)
\SetWidth{1.5}
\Line(222,70)(230,80)
\SetWidth{.5}
\DashLine(222,30)(222,70){2}
\Text(232,54)[]{$H$}
\Text(210,70)[]{$\ell_j$}
 \Text(210,30)[]{$N_\beta$}
\Text(237,85)[]{$N_\alpha$}
 \Text(237,20)[]{$\ell_i$}
\Text(315,0)[]{$(c)$}
\ArrowLine(265,80)(290,50)
\ArrowLine(265,20)(290,50)
\Text(275,85)[]{$Q$}
\Text(275,20)[]{$\bar t$}
\DashLine(290,50)(340,50){2}
\Text(310,60)[]{$H$}
\ArrowLine(340,50)(365,20)
\SetWidth{1.5}
\Line(340,50)(348,59.6)
\Line(360,74)(365,80)
\SetWidth{.5}
\DashCArc(354,66.8)(9.4,60,240){2}
\ArrowArc(354,66.8)(9.4,-120,60)
\Text(337,60)[]{$N_\beta$}
\Text(343,82)[]{$H,\bar H$}
\Text(379,60)[]{$\ell_j,\bar \ell_j$}
\Text(372,85)[]{$N_\alpha$}
\Text(372,20)[]{$\ell_i$}
\end{picture}
\end{center}
\caption{\small 
Diagrams contributing to the CP asymmetry 
in $Q\bar t \leftrightarrow N_\alpha \ell_i$ scatterings. 
}
\label{fig1}
\end{figure}
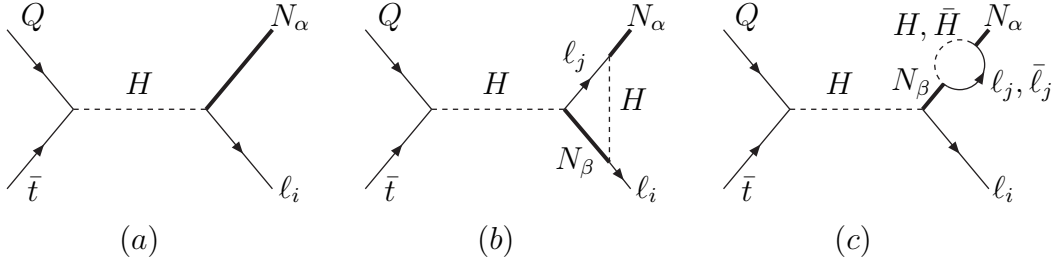

The vertex contribution to the CP asymmetry in the scattering $Q\bar t
\leftrightarrow N_\alpha \ell_i$ with the Higgs exchanged in the $s$-channel 
 arises from the interference between the diagrams in fig.~\ref{fig1}$(a)$
and \ref{fig1}$(b)$.  Expressed in terms of the CP difference between the
squared invariant amplitudes $|\mathcal{M}|^2- |\bar \mathcal{M}|^2 $
this contribution reads
\begin{eqnarray}
\left[|\mathcal{M}(Q  \bar{t} \rightarrow N_\alpha
  \ell_i)|^2-|\bar{\mathcal{M}}|^2\right] (vertex) =- \frac{6}{ \pi}
\frac{ p_1\cdot p_2\ p'_1\cdot p'_2 }{s^2}
  |h_t|^2 \times \qquad \cr
 \sum_\beta {\rm Im}[\lambda_{i\alpha} \lambda^*_{i \beta}(\lambda^\dag 
\lambda)_{\beta
  \alpha}] \frac{M_\alpha M_\beta}{s-M_\alpha^2} \left\{ 
\left[ 1 -
  \frac{M_\alpha^2 + M_\beta^2 - s}{M_\alpha^2 - s} \ln \left(\frac{|M_\alpha^2
    + M_\beta^2 - s|}{M_\beta^2} \right) \right] - \right. \cr
   \left. \theta(s-M_\beta^2) \left[ \frac{s-M_\beta^2}{s} + \frac{M_\alpha^2 +
  M_\beta^2-s}{s - M_\alpha^2} \ln \left(\frac{s |M_\alpha^2 +
  M_\beta^2-s|}{M_\beta^2 M_\alpha^2} \right)\right]\right\} .
\end{eqnarray}
Here $\theta$ is the step function ($\theta(x)=1$ for $x>0$ and $0$
otherwise), $s$ is the squared center of mass energy, $p_1, p_2, p'_1$ and
$p'_2$ are the momenta of $t,Q ,\ell_i$ and $N_\alpha$ respectively, and
$M_\beta$ is the mass of $N_\beta$ (note that for $\beta=\alpha$ there is no
contribution since the product of the relevant Yukawa couplings is real).  The
overall factor of 6 corresponds to the summation over the gauge degrees of
freedom.  Note that the term proportional to $\theta(s-M_\beta^2)$ appears due
to the possibility of performing a new cut in the one-loop graph in
fig.~\ref{fig1}$(b)$ involving the $N_\beta$ and lepton lines.  Since
$N_\beta$ can go on-shell only when the center of mass energy is sufficiently
large ($s> M^2_\beta$) this contribution is relevant only for temperatures not
much smaller than $M_\beta$.  Our results hold in the zero temperature limit;
in particular we take all the particles, except the heavy Majorana neutrinos,
to be massless.  At high temperatures, finite temperature effects can induce
non negligible corrections to our expressions. In particular, when $M_{1} <
M_H(T)+M_{\ell_i}(T)$ (implying that decays and inverse decays are blocked
\cite{gi04}) we expect that thermal masses will also have the effect of
suppressing the scattering CP violating asymmetries.
%
%
%
%
%

Due to crossing symmetry, the CP asymmetry for the processes $N_\alpha \bar{t}
\rightarrow \bar{Q } \ell_i$ and $N_\alpha Q \rightarrow t \ell_i$ in which
the Higgs is exchanged in the $t$-channel can be obtained from the previous
result by replacing the Mandelstam variable $s$ by $t$.  Note that for
massless quarks $t \leq 0$ so that $\theta(t-M_\beta^2)=0$ and hence in this
case no new cut is present.  

The cross sections are obtained by integrating the modulus squared of the
invariant amplitudes:
\begin{equation}
\sigma= \frac{1}{64 \pi^2 (E_1+E_2)^2}\int 
\frac{|\bf{p^\prime_1}|}{|\bf{p_1}|} |\mathcal{M}|^2 {\rm d} \Omega^\prime_1,
\end{equation}
where $p_i=(E_i,\bf{p_i})$, $p^\prime_i=(E_i^\prime,\bf{p^\prime_i})$ $(i=1,2)$ are
the momenta of the initial and final particles respectively, and the cross
section for the CP conjugate process $\bar \sigma$ is defined in the same way.
In terms of the cross sections, the CP asymmetry of the $s$-channel
scattering is:
\begin{eqnarray}
\left[\sigma(Q  \bar{t} \rightarrow N_\alpha \ell_i)-
 \bar{\sigma}\right](vertex) =  -\frac{1}{8 \pi} \frac{\sum_j [\sigma(Q  \bar{t} \rightarrow N_\alpha
 \ell_j)+\bar{\sigma}]}{(\lambda^\dag\lambda)_{\alpha\alpha}}\times \qquad \cr
\sum_\beta
  {\rm Im}[\lambda_{i\alpha} \lambda^*_{i \beta}(\lambda^\dag 
\lambda)_{\beta
  \alpha}] \frac{M_\alpha M_\beta}{s-M_\alpha^2}
 \left\{ \left[ 1 -
  \frac{M_\alpha^2 + M_\beta^2 - s}{M_\alpha^2 - s} \ln \left(\frac{|M_\alpha^2
    + M_\beta^2 - s|}{M_\beta^2} \right) \right] - \right.  \cr
   \left. \theta(s-M_\beta^2) \left[ \frac{s-M_\beta^2}{s} + \frac{M_\alpha^2 +
  M_\beta^2-s}{s - M_\alpha^2} \ln \left(\frac{s |M_\alpha^2 +
  M_\beta^2-s|}{M_\beta^2 M_\alpha^2} \right)\right]\right\} .
\label{asyvert}
\end{eqnarray}
Unlike what happens in the case of decays, the asymmetry $\sigma -\bar\sigma$
of the scattering rates now depends on $s$, so that the convolution necessary
to obtain the asymmetry $\Delta\gamma^{Q\bar t}_{N\ell}$ of the thermally
averaged rates doesn't lead to a simple analytical expression, and has to be
performed numerically.

For the CP asymmetry in the $t$-channel scattering there is no simple
analytical expression.  Also, the usual infrared divergence appears in the
limit $t\to 0$, which can be regularized by replacing the factor $1/t$ coming
from the (massless) Higgs propagator by $1/(t-m_H^2)$, and using here the
Higgs thermal mass. 

The CP asymmetry in scatterings coming from the wave function piece
(interference between the diagrams in figs.~\ref{fig1}$(a)$ and
\ref{fig1}$(c)$) turns out to be always the same as the CP asymmetry for the
decays:
\begin{equation}
\frac{\Delta \gamma^{Q  \bar{t}}_{N_\alpha \ell_i}(wave)}{\sum_j
 ( \gamma^{Q  \bar{t}}_{N_\alpha \ell_j}+ \gamma^{\bar Q  t}_{N_\alpha
 \bar\ell_j})}  = 
\frac{\Delta  \gamma^{N_\alpha}_{\ell_i H}(wave)}{\sum_j
(  \gamma^{N_\alpha}_{ \ell_j H}+  \gamma^{N_\alpha}_{ \bar\ell_j \bar H}
)} \equiv \varepsilon^i_\alpha(wave),
\end{equation}
where  the decay asymmetry is defined in the usual way as
$\varepsilon^i_\alpha\equiv \Delta\gamma^{N_\alpha}_{\ell_i
  H}/\gamma_D^{N_\alpha \to 2}$.

\begin{figure}[t!]
\centerline{\protect\hbox{
\epsfig{file=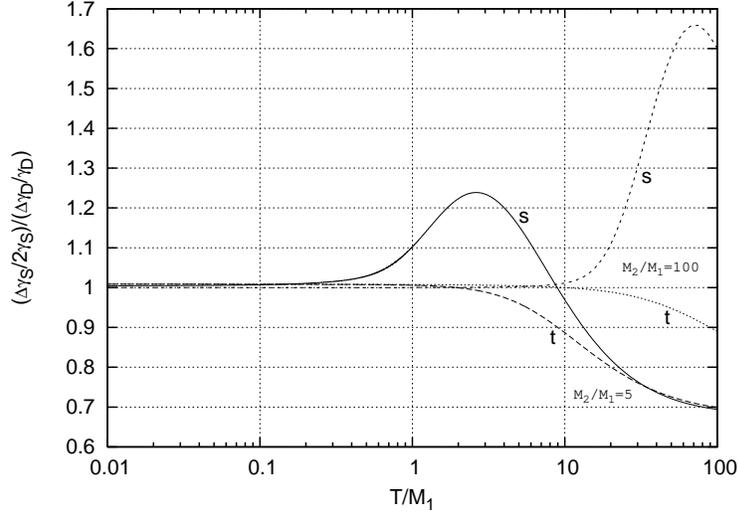 ,width=7cm,angle=270}}}
\caption{\small 
  Ratio between the CP violating asymmetry in scatterings $\Delta\gamma^{Q\bar
    t}_{N\ell_i}/2\gamma^{Q\bar t}_{N\ell_i}$ and the corresponding quantity
  for decays $\Delta\gamma^N_{\ell_iH}/2 \gamma^N_{\ell_iH}$, vs. $T/M_1$ for
  two ratios between the masses of the two lightest heavy neutrinos
  $M_2/M_1=5$ and $M_2/M_1=100$.  The contributions from $s$ and
  $t$-channel Higgs exchange processes are shown separately.}
\label{ratio.fig}
\end{figure}

The ratio between the CP violating scattering asymmetries and their
approximate expressions in terms of the asymmetry in decays derived in
ref.~\cite{aba06b}, are shown in figure~\ref{ratio.fig} as a function of
$T/M_1$\footnote{For definiteness, we neglected in this plot the contribution
  to the wave part proportional to
  $M_\alpha(\lambda^\dagger\lambda)_{\alpha\beta}$, keeping the one
  proportional to $M_\beta(\lambda^\dagger\lambda)_{\beta\alpha}$, in which
  case vertex and wave contributions become proportional to the same
  combination of couplings, see ref.~\cite{co96}.}.  The results for processes
with the Higgs exchanged in the $s$ and in the $t$-channels are presented
separately.  Two illustrative values for the ratio of the two lightest
Majorana neutrino masses have been used: $M_2/M_1=5$ and $M_2/M_1=100$.  The
effects of $N_3$ have been ignored (this would correspond either to the case
$M_3\gg M_2$ or to the situation in which the complex phase in the combination
of Yukawas associated to $N_3$ is particularly suppressed).  It is apparent
that this ratio starts to deviate from unity already for $T>M_2/10$ and that
deviations of few tens of percent can appear for $T$ approaching $M_2$.  Since
the relevant temperature for leptogenesis in these scenarios is typically $0.1
<T/M_1<10 $, the approximations adopted in ref.~\cite{pi04-pi05,aba06b} should
be good if $M_2/M_1\gg 10$, while some corrections appear for milder
hierarchies at temperatures $T>M_2/10$. It is also easy to show analytically,
starting from eq.~(\ref{asyvert}), that a factorization for the vertex part,
analogous to that of the wave part, is obtained in the limit of $M_1$ and
$T\ll M_2$.

Regarding the scattering processes with gauge bosons, such as $N\ell\to\bar
HA$, $NH\to\bar \ell A$  or $NA\to \ell H$, the associated CP asymmetries can
be obtained in a similar way, computing the interferences between tree and one
loop scattering amplitudes. One significant difference is that now box
diagrams are present, in which the gauge boson is attached to a lepton or
Higgs in the loop of the vertex like diagrams, leading to more
complicated expressions. The absorptive parts can be obtained using
Cutkowski's rule and  new cuts appear, but again only for $s> M_\beta^2$, so
that in the hierarchical limit the factorized expression holds for $T\ll M_2$,
as was the case in the scatterings with quarks discussed before.

\begin{figure}[ht!]
\centerline{\protect\hbox{
\epsfig{file=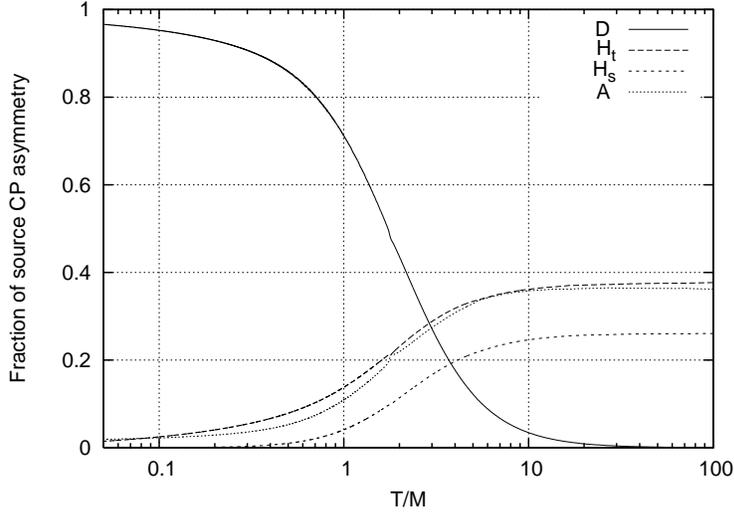 ,width=7cm,angle=270}}}
\caption{\small Fractional contribution to the source terms from the CP
  asymmetries in decays ($D$) and in scatterings ($s$ and $t$ channel Higgs
  exchange and scatterings involving gauge bosons) for $M_2\gg M_1$, 
 obtained in the zero temperature approximation. }
\label{fract5}
\end{figure}

The fractional contributions to the source terms in the BE for $Y_{\Delta_i}$
arising from decays and scatterings are shown in fig.~\ref{fract5} using the
factorized expressions in terms of the decay asymmetries.  We present
separately the results for two body decays $F_D\equiv
\Delta\gamma^N_{\ell_iH}/\sum\Delta\gamma$, $s$-channel $F_{H_s}\equiv
\Delta\gamma^{Q\bar t}_{\ell_iN}/\sum\Delta\gamma$, $t$-channel Higgs exchange
processes $F_{H_t}\equiv 2\Delta\gamma^{NQ}_{\ell_it}/\sum\Delta\gamma$ and
the gauge boson contribution\footnote{The contribution of gauge boson
  scatterings has been estimated using the expressions given in \cite{gi04}.}
$F_A\equiv (\Delta\gamma^{NA}_{\ell_iH}+\Delta\gamma^{N\bar H}_{\ell_iA}+
\Delta\gamma^{N\bar\ell_i}_{AH})/\sum\Delta\gamma$ where
\begin{equation}
\sum\Delta\gamma\equiv \Delta\gamma^N_{\ell_iH}+\Delta\gamma^{Q\bar
  t}_{\ell_iN}+ 2\Delta\gamma^{NQ}_{\ell_it}+\Delta\gamma^{NA}_{\ell_iH}+
\Delta\gamma^{N\bar  H}_{\ell_iA}+ \Delta\gamma^{N\bar\ell_i}_{AH},  
\end{equation}
 while the contributions from decays into three body final states are always
negligible and are not shown (see the appendix).  These fractions are
independent of the value of the neutrino Yukawa couplings adopted and, in the
hierarchical limit in which the factorization is valid, they are also
independent of the value of $M_2/M_1$.   From fig.~\ref{fract5} it is seen
that scattering CP asymmetries are the dominant source term for $T>2M_1$, and
hence can play an important role in the early leptogenesis
phase.\footnote{When thermal masses are taken into account, at very high
  temperatures ($T\gsim 7 M_{1}$) the condition $M_H(T) >
  M_{1}+M_{\ell_i}(T)$ is met and the decay $H\to N_1\ell_i$ can occur.
  Since the asymmetry for this decay has a large enhancement from thermal
  effects \cite{gi04}, in this temperature regime actually the 
Higgs decays would become the
  dominant source of the lepton asymmetry.}

\section{Results}

\begin{figure}[ht!]
\centerline{\protect\hbox{
\epsfig{file=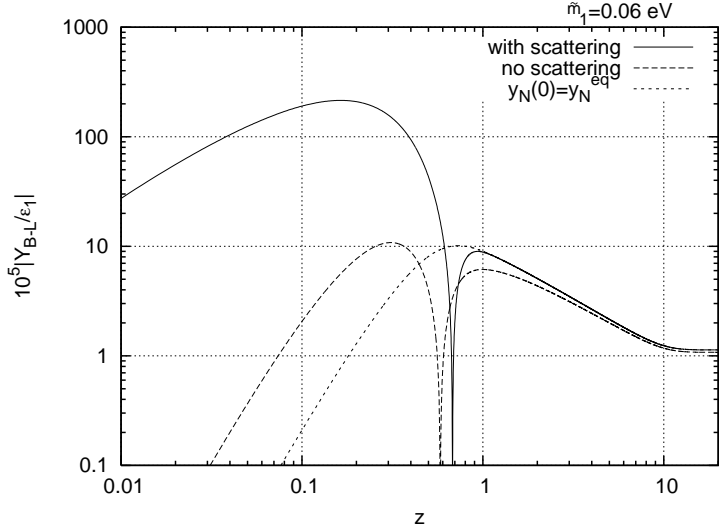 ,width=7cm,angle=270}}}
\caption{\small Evolution of the $B-L$ asymmetry as a function of $z=M_1/T$
  with the CP asymmetries in scatterings neglected (dashed line) and included
  (solid lines). The scattering asymmetries are approximated by using the
  asymmetry in decays. This figure corresponds to a strong washout regime 
  with $\tilde m_1=6\times 10^{-2}$~eV. For comparison, the value that would
  be obtained starting with an equilibrium $N$ density is also displayed
  (dotted line).  }
\label{figure5}
\end{figure}

In fig.~\ref{figure5} we display the results of the integration of the BE
adopting $M_1=10^{11}$~GeV and $\tilde m_1\equiv v^2(\lambda^\dagger
\lambda)_{11}/M_1=0.06$~eV (where $v$ is the Higgs VEV).  To avoid
complications with $\tau$-flavor effects that are active in this temperature
regime, we also assume a flavor `aligned' situation in which the lepton
doublet $\ell_1$ to which $N_1$ decays has no $\tau$ flavor component, that is
$K_\tau\equiv |\langle \ell_1|\ell_\tau\rangle |^2=0$ (and similarly for $\bar
\ell_1$) .  The dashed line corresponds to the case in which the contributions
of the CP scattering asymmetries to the source term are ignored (see
ref.~\cite{na05} for more details), while the solid line depict the results
obtained by including the CP asymmetries of the scattering processes, adopting
their factorized expression (using the exact expressions would slightly modify
the asymmetries for $T>M_1$, but the final values would be almost unchanged).
The dotted line is the asymmetry that would result had the initial density of
$N$ be the equilibrium one.  It is apparent that for $T>M_1$ the scattering
processes have a  large effect in the production of the lepton asymmetry.
This example corresponds to a case of strong washout, with $\tilde m_1\gg
10^{-3}$~eV. In particular, we see that in this case the washouts affect the
evolution of the lepton asymmetries up to $z\simeq 10$. Hence, even if at
early times large additional sources of CP violation are present, late
washouts turn out to be decisive in determining the final asymmetry, which
ends up being equal to the one that would be obtained had one started with
$Y_N=Y_N^{eq}$.  This also means that in this regime the final asymmetry
becomes essentially independent from the conditions at early times.

 We recall here that a crucial point in thermal leptogenesis is that if the
washout processes were switched off completely during the whole leptogenesis
phase (and if the dependence of the CP asymmetries on the temperature that is
induced by thermal effects was also ignored) the inclusion in the BE of the
sources of CP violation from scatterings would yield a zero final asymmetry
\cite{aba06b}.  This can be seen by writing the different $\Delta L_i =1$
scattering CP asymmetries in terms of their approximate expressions, i.e.
\begin{equation}
\Delta\gamma^{NX}_Y\simeq \frac{\gamma^{NX}_Y}{\gamma^N_{\ell_i H}}\Delta
\gamma^N_{\ell_iH}.  
\end{equation}
Then, the source term in the BE for $Y_{\Delta_i}$ can be rewritten just in
terms of the decay asymmetry $\Delta \gamma^N_{\ell_iH}$ as 
\begin{equation}
(\dot{Y}_{\Delta_i})_s\simeq -(y_N-1)\frac{\Delta
\gamma^N_{\ell_iH}}{\gamma^N_{\ell_iH}}   \sum_{\Delta L_i=1}
\gamma^{NX}_Y.
\label{ypsrc3a}
\end{equation}
By using the relation 
\begin{equation}
\frac{\gamma^N_{\ell_iH}}{\sum_j(\gamma^N_{\ell_jH}+\gamma^N_{\bar\ell_j\bar
    H})} =\frac{\sum_{\Delta L_i=1}
\gamma^{NX}_Y}{\sum_{X,Y}\gamma^{NX}_Y},  
\end{equation}
where the sum in the denominator in the r.h.s. is over all the processes
$\gamma^{NX}_Y$ for which $|\Delta L_j| = 1$ and over all the flavors $j$, and
defining the flavor asymmetry $\varepsilon_1^i=\Delta\gamma^N_{\ell_i
  H}/\sum_j (\gamma^N_{\ell_jH}+\gamma^N_{\bar\ell_j\bar H})$ we can finally
rewrite the source term as
\begin{equation}
(\dot{Y}_{\Delta_i})_s\simeq -(y_N-1)\,\varepsilon^i_1 \sum_{X,Y}\gamma^{NX}_Y. 
\label{ypsrc3b}
\end{equation}
Combining now the above expression with the BE for $Y_N$ we  obtain
\begin{equation}
\dot{Y}_N-\frac{\dot{Y}_{\Delta_i}}{\varepsilon^i_1}\simeq -
\frac{(\dot{Y}_{\Delta_i})^w}{\varepsilon^i_1}.
\end{equation}
In the absence of washouts the r.h.s. of the equation above would then
vanish, 
and in the approximation in which $\varepsilon^i_1$ is taken as independent of
the temperature, the quantity $Y_N-Y_{\Delta_i}/\varepsilon^i_1$ would hence
be constant. Then, for thermal leptogenesis scenarios in which the $N$ density
and the lepton asymmetries vanish initially, this quantity will just be zero,
showing that the $Y_{\Delta_i}$ asymmetries generated at early times are
erased at later times as the $N_1$ disappear by decays or scatterings.  Then,
as was pointed out in \cite{gi04}, any effect that breaks this cancellation,
as for example a dependence of the CP asymmetries on the temperature, could be
numerically important.  The cancellation will no longer hold also if some
washout processes are particularly efficient at temperatures $T>T_{0}$, where
$T_0$ is the temperature at which the lepton asymmetry changes sign (note that
this type of washouts could yield an enhancement in the final asymmetry, while
in general late washouts at $T<T_0$ always tend to reduce its final value). Of
course, in the cases when the heavy $N$ states are produced through other
processes not related to the ones giving rise to the CP asymmetries, such as
via scatterings involving heavy right-handed $W$ or additional $Z'$ bosons, or
are produced non-thermally via e.g.  inflaton decays, the leptogenesis initial
condition $Y_N(0)\neq 0$ would directly prevent the cancellation of the lepton
asymmetry.

In other words, the origin of the cancellation can be understood as follows.
At any time there are three kinds of possible sources for the lepton
asymmetries: off-shell scatterings, processes producing real $N$s and
processes in which $N$s are destroyed. In general, the lepton asymmetry
produced by the off-shell scatterings is twice as large, and with opposite
sign, as that associated to processes in which real $N$s are produced, but
that associated to processes in which real $N$s are destroyed depends on the
$N$ density $Y_N$. If $Y_N$ equals the equilibrium density $Y_N^{eq}$, the
asymmetry of $N$ production and destruction processes are equal, and hence the
sum of the three asymmetries cancel. On the other hand, if $Y_N<Y_N^{eq}$ the
asymmetry produced by off-shell processes dominates, while the opposite
happens if $Y_N>Y_N^{eq}$. This is why the sources of lepton asymmetry in the
BE are just proportional to $Y_N-Y_N^{eq}$. Now, when the factorization of the
scattering asymmetries in terms of the decay asymmetry holds, one finds that
the sources of the lepton asymmetries are just $\epsilon_1$ times the sources
of the $N$ density (assuming that no other processes besides the Yukawa
couplings produce $N$s). This means that if $\epsilon_1$ is constant ($T$
independent) and we ignore the washout processes, the total integrated change
in the lepton asymmetry will be just $\epsilon_1$ times the total change in
$Y_N$. Hence, if this last vanishes, as is the case when the initial condition
is that of vanishing $N$ density, the final leptonic density would also
vanish.

\begin{figure}[ht!]
\centerline{\protect\hbox{
\epsfig{file=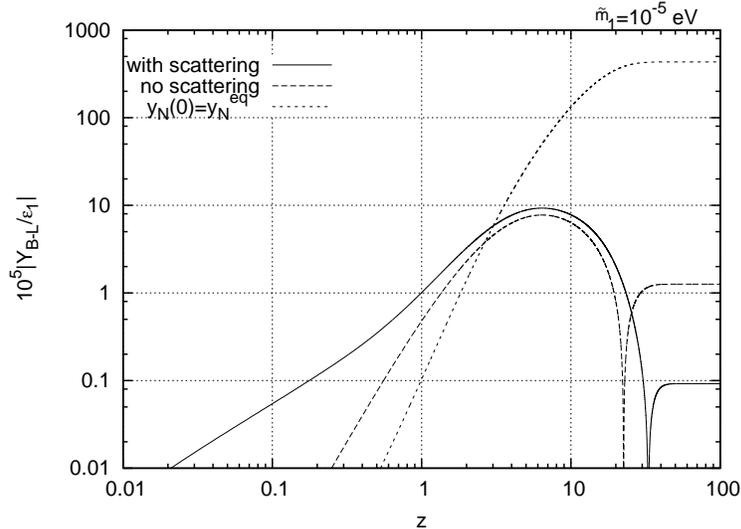 ,width=7cm,angle=270}}}
\caption{\small 
Same as Figure \ref{figure5} but in a regime of weak washout, with $\tilde
m_1=10^{-5}$~eV.  }
\label{figure6}
\end{figure} 

 Clearly, the impact of including CP scattering asymmetries in the BE is
qualitatively different in the strong and in the weak washout regimes.  In the
weak washout regimes (corresponding to values of $\tilde m_1<10^{-3}$~eV) the
effect of late washouts is negligible, the asymmetry is strongly affected by
the cancellation and thus its final value turns out to be rather sensitive to
the amount of washouts in the early phases.  This is illustrated in the
example in fig.~\ref{figure6}, that corresponds to the value $\tilde
m_1=10^{-5}$~eV. In this figure we compare the evolution of the asymmetries in
the two cases when the CP asymmetries of the processes involving the
top-Yukawa and the gauge interactions are included or are left out (all
scatterings are in any case included as sources for $N$ production).  It is
apparent that when the sources of CP violation from scatterings are included
the effects of the cancellation strongly reduce the final asymmetry obtained.
Since only weak washouts are present, the cancellation remains quite effective
and the final value of the asymmetry is rather small.  In this regime, the
final asymmetries are also much smaller than the asymmetries that would result
starting with an equilibrium density for the $N$'s
 since in that case no asymmetry is generated at early times, and no
 cancellation can occur.

\begin{figure}[ht!]
\centerline{\protect\hbox{
\epsfig{file=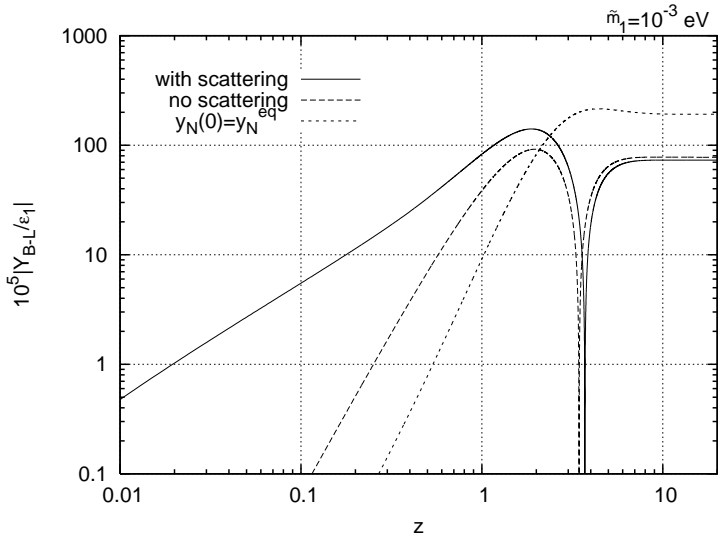 ,width=7cm,angle=270}}}
\caption{\small 
Same as Figure \ref{figure5} but in a regime of intermediate washout, 
with $\tilde m_1=10^{-3}$~eV.  }
\label{figure7}
\end{figure} 

 Finally, fig.~\ref{figure7} gives an example with an intermediate washout
strength $\tilde m_1=10^{-3}$~eV.  In this case an intermediate behavior
between those of the weak and strong washouts is observed: similarly to the
strong washout case, the final asymmetry remains almost unchanged whether the
CP scattering asymmetries are included or not. However, as in the case of weak
washouts, its value remains well below what would be obtained by starting with
an equilibrium density of $N$'s.

\section{Conclusions}

In this paper we have computed the CP asymmetries of scattering processes
involving the top quark and the gauge bosons.  CP violation in scatterings
gives an important contribution to the generation of a lepton asymmetry at
high temperatures ($T\gsim 2M_1$), and in particular in the zero temperature
approximation adopted in our calculations this contribution is by far the
dominant one.  We have compared our results with the approximate expressions
of the scattering CP asymmetries in terms of the decay asymmetry, concluding
that in scenarios in which the heavy Majorana neutrino masses are sufficiently
hierarchical this approximation provides reasonably accurate results.  We have
shown that when the sources of CP violation in scatterings are included in the
BE, in the limit of very weak washouts a strong cancellation between the
asymmetries generated at early times
 (when $Y_N<Y_N^{eq}$), 
and the asymmetry of opposite sign generated at later times
 (when $Y_N>Y_N^{eq}$) 
takes place, sizably suppressing the final lepton asymmetry with respect to
the cases in which CP asymmetries in scatterings are neglected.

In the strong washout regimes, for which the final lepton asymmetry is
almost independent of the conditions at early times, such as  the initial 
value of the right-handed neutrino density, the final results are  instead 
essentially  unaffected by the
inclusion of the new sources of CP violation.

\section*{Acknowledgments}
We are grateful to Y. Nir for useful discussions.  The work of E.N. is
supported in part by the Istituto Nazionale di Fisica Nucleare (INFN) in
Italy, and by Colciencias in Colombia under contract 1115-333-18739.  The work
of E. R. is partially supported by a PICT grant from ANPCyT.

\section{Appendix: Three body $N$ decay}
The three body decay width for $N_1\to \ell_i \overline{Q}_3 t$ is
\begin{eqnarray}
\label{tridec}
\nonumber
\Gamma(N_1 \rightarrow \ell_i \overline{Q}_3 t)&=\frac{3}{16\pi^2}
h_t^2  \Gamma_0(N_1 \rightarrow \ell_i H)
\int_{(\sqrt{a_t}+\sqrt{a_Q})^2}^{(1-\sqrt{a_\ell})^2} {\rm d}x \,
\frac{(x-a_t-a_Q)(1+a_\ell-x)}{(x-a_H)^2+a_Hc_H}\\
&\quad \frac{1}{x} \left( \left[ (x-a_Q+a_t)^2-4xa_t \right] \left[
    (1-x-a_\ell)^2-4xa_\ell \right]\right)^{1/2},
\end{eqnarray}
where $a_y\equiv(m_y/M_1)^2$ and $c_H\equiv (\Gamma_H/M_1)^2$, with $\Gamma_H$
being the decay width of the Higgs boson. $\Gamma_0 (N_1 \rightarrow \ell_i
H)=|\lambda_{i1}|^2\, M_1 / 16\pi$ is the two body decay width of $N_1$ (at
zero temperature) and the integration variable is $x=(p_t + p_Q)^2/M_1^2$,
where $p_t$ and $p_Q$ are the momenta of $t$ and $\overline{Q}_3$
respectively.

We assume $m_t+m_{Q_3}>m_H$, which is generally valid at high $T$ if thermal
masses are considered and also at $T=0$ if the Higgs boson is not too heavy.
In this case the Higgs boson exchanged in the internal line of the three body
decay cannot be on-shell, and the Higgs width (parametrised by $c_H$) can be
neglected.  (Note that a resonant contribution would in any case correspond to
the two body decay $N_1\to \ell_i H$ rather than to a genuine three body
process.)  In the zero temperature limit $a_t,\ a_Q,\ a_H\to 0$ the integral
in eq.~(\ref{tridec}) would get a large enhancement from the region
corresponding to small values of $x$.  However, for finite values of the
thermal masses this enhancement is not present, and in particular for
$T/M_1>10^{-2}$ the three body decay rate is always less than 6\% of the two
body decay rate.  Note also that, due to the effects of thermal masses, the
phase space for both decays actually gets closed when T approaches $M_1$
\cite{gi04}.

\end{document}